\documentclass[aps,pra,groupedaddress,twocolumn,floatfix]{revtex4}

\usepackage[dvips]{graphicx}
\usepackage{dcolumn}
\usepackage{bm} 
\usepackage{amsmath,amsthm,amssymb}
\usepackage{graphicx}
\usepackage{epstopdf}
\usepackage{color}
\usepackage{siunitx}
\usepackage{tabularx} 
\usepackage{booktabs} 
\DeclareSIUnit\gauss{G}
\DeclareSIUnit\atoms{atoms}
\DeclareSIUnit\grav{g}
\DeclareSIUnit\efield{(\volt/\centi\meter)}
\DeclareSIUnit\debye{Debye}
\usepackage{xfrac}

\usepackage{units}				
\usepackage{natbib}
\begin{document}
\title{Designs of magnetic atom-trap lattices for \mbox{quantum} simulation experiments } 

\author{A.\ L.\ La Rooij\footnote{Present address: Department of Physics, University of Strathclyde, 107 Rottenrow East, G4 0NG Glasgow, United Kingdom}, H.\ B.\ van Linden van den Heuvell,   R.\ J.\ C.\ Spreeuw}  

\affiliation{Van der Waals-Zeeman Institute, Institute of Physics, University of
Amsterdam, P.O. Box 94485, 1090 GL, Amsterdam, The Netherlands.}
\date{\today}

\begin{abstract}
We have designed and realized magnetic trapping geometries for ultracold atoms based on permanent magnetic films. Magnetic chip based experiments give a high level of control over trap barriers and geometric boundaries in a compact experimental setup. These structures can be used to study quantum spin physics in a wide range of energies and length scales. By introducing defects into a triangular lattice, kagome and hexagonal lattice structures can be created.   Rectangular lattices and (quasi-)one-dimensional structures such as ladders and diamond chain trapping potentials have also been created.  Quantum spin models can be studied in all these geometries with Rydberg atoms, which allow for controlled interactions over several micrometers. We also present some nonperiodic geometries where the length scales of the traps are varied over a wide range. These tapered structures offer another way to transport large numbers of atoms adiabatically into subwavelength traps and back.  \\
\end{abstract}
\maketitle

\section{Introduction}
Experiments with cold atoms trapped in lattice-type trapping potentials have opened up new windows on condensed matter phenomena 
\cite{gross2017quantum}. The unprecedented control over many-body quantum systems that these quantum simulators have to offer allows experimentalists to study increasingly complex systems. It is the hope of many that by these means it will be possible to emulate some of the outstanding challenges, such as  high-$T_{C}$ superconductivity \cite{hofstetter2002high}, frustrated magnetism \cite{struck2011quantum}, and even high energy gauge theories like QCD \cite{banerjee2013atomic}. Most results of the past decades have been obtained in optical lattice experiments where lattices of trapped atoms are created by standing waves of laser light. These experiments are mostly focused on lattices with open or harmonically confined boundaries and with a single lattice type. The method of chip based magnetic trapping provides an alternative to optical lattices where one has more freedom in the construction of trapping geometries and length scales. 

In recent years, magnetic lattices with \unit[10]{$\mu$m} lattice spacing have been created to trap atomic ensembles 
\cite{jose2014periodic,leung2014magnetic}. In these lattices Rydberg interactions could be exploited to control interactions between the sites which are necessary for most  quantum simulation and quantum information experiments \cite{saffman2010quantum, wilk2010entanglement, whitlock2017simulating,zeiher2016many}. 
It was recently shown theoretically that Rydberg atoms can indeed be trapped in tight magnetic traps and even magic trapping conditions may be achieved \cite{boetes2018trapping}. Note, however, that experiments on Rydberg excitation near surfaces have so far been hampered by stray electric fields emanating from adatoms \cite{tauschinsky2010spatially, naber2016adsorbate}; see also Sec. II. 
Experiments so far have employed all triangular, square, and oblique geometries in two dimensions \cite{leung2014magnetic} and elongated one-dimensional lattices \cite{jose2014periodic}. 

Here we present several geometries to trap ultracold atoms, which will allow many new quantum simulation studies. First, in Sec. \ref{defSection} we explain the methods used in the designs.  Kagome and hexagonal (honeycomb) structures will be presented in Sec. \ref{triHoneyKagome}. These geometries are particularly important for studies of frustration and quantum magnetism in two dimensions \cite{balents2010spin}. In Sec. \ref{ladders} ladders and diamond chains are introduced which may be used to study magnetic phases of spin models in atomic chains. With elements that interrupt the periodicity of a lattice we are able to create barriers along orthogonal directions in the plane. These can be used to isolate plaquettes of a finite number of traps. Finally, in Sec. \ref{tapersec} a tapered lattice is presented wherein a varying lattice spacing yields tapered structures which provide a natural bridge between traps at the Rydberg scale and nanoscale magnetic lattices which have recently been introduced \cite{la2018deposition, herrera2015sub, wang2017trapping}. The extra applications that these barriers and tapered structures have on subwavelength magnetic traps will be discussed in a separate paper. 

\section{Magnetic fields of patterned films} \label{defSection}

We assume that the  spin of a moving atom follows the local direction of the magnetic field $\boldsymbol{B}(\boldsymbol{r})$ adiabatically. The magnetic potential energy $-\boldsymbol{\mu} \cdot \boldsymbol{B}(\boldsymbol{r})$ is then proportional to the magnitude of the field, $U(\boldsymbol{r})=\mu_B m_F g_F B(\boldsymbol{r})$. Here $\mu_B$ is the Bohr magneton, $m_F$ is the magnetic quantum number, and $g_F$ the Land\'e factor for an atomic level with total angular momentum $F$. Atoms in a ``low-field-seeking" state (with $m_F g_F>0$) can thus be trapped in a local  magnetic field minimum.  Majorana losses to nontrapped states can be neglected as long as the trap frequencies remain much smaller than the Larmor frequency  $\omega_{\mathrm{L}} = \mu_{B} g_{F} \lvert B_{\mathrm{bot}} \lvert / \hbar  $ \cite{sukumar1997spin}. Here $B_{\mathrm{bot}}$ is the absolute value of the magnetic field at the trap bottom.  

The building block for all lattice geometries below is the Ioffe-Prichard trap (IPT) \cite{pritchard1983cooling, gott1962nuclear}, which is a widely used technique to create magnetic field minima. As a film with out-of-plane magnetization can be described by an equivalent edge current $I_{M}$ which runs along the edge of the film, both  current carrying wires and patterned permanent magnets can be used to trap atoms.

In this paper we are concerned with designing magnetic trapping potentials generated by permanently magnetized films with a thickness $h\lesssim $ \unit[300]{nm}. The distance to the chip surface is typically large compared to the film thickness,  $z\gg h$, so we neglect the finite film thickness. We describe structures made out of perpendicularly magnetized films, patterned in a binary fashion; i.e., we assume that in any given location on the chip we have either the full film thickness or no magnetic material at all. Instead of the bulk magnetization $M$ we use the two-dimensional magnetization, i.e., the magnetic dipole per unit area: $I_{M}= Mh$. Our recently constructed experiment houses a chip with a \unit[50]{nm} thick film of magnetized FePt with a magnetization of $M=$ \unit[800]{kA/m} \cite{la2018deposition}. Because several structures that we will present have been realized on this chip, we will consider this thickness and equivalent edge current $I_{M}= $ \unit[0.04]{A}. 

As a basis for the different trapping geometries that we will develop, we use the triangular and square lattice patterns that have been developed by Fourier space optimization \cite{schmied2010optimized}. For most of the presented structures it is convenient to express the fields in terms of Fourier series. Furthermore, in the region of space above the chip surface, the static magnetic field can be written as the gradient of a scalar potential, $\boldsymbol{B}(\boldsymbol{r})=-\nabla\Phi_M(\boldsymbol{r})$. 

Neglecting the finite film thickness, we define  $M_2$ as the surface density of magnetic moment. 
Taking this two-dimensional  magnetization to be periodic, $M_2(\boldsymbol{\rho})=M_2(\boldsymbol{\rho}+\boldsymbol{r_{1}})=M_2(\boldsymbol{\rho}+\boldsymbol{r_{2}})$, with 
$\boldsymbol{r_{i}} $ the basis vectors of the lattice and $\boldsymbol{\rho}$ the in-plane two-dimensional coordinate,
the magnetization can be written as a two-dimensional Fourier series, 
\begin{eqnarray}  \label{eq:phiBcos}
  M_2(\boldsymbol{\rho})= I_{M} \sum_{m,n=-\infty}^{\infty} 
   & & \hspace{-1em} C_{nm} \cos[(n \boldsymbol{K}_1+m \boldsymbol{K}_2)\cdot\boldsymbol{\rho} ] + \nonumber  \\
  & & \hspace{-1em} S_{nm} \sin[(n \boldsymbol{K}_1+m \boldsymbol{K}_2)\cdot\boldsymbol{\rho} ] 
\end{eqnarray}   
such that $ M_2(\boldsymbol{\rho}) / I_M  $ is $0$ or $1$. The vectors $\boldsymbol{K}_1,\boldsymbol{K}_2$ are the basis vectors of the reciprocal lattice, defined by $ \boldsymbol{K_{i}} \times \boldsymbol{r_{j}}  = 2 \pi \delta_{ij} $, and the area of the unit cell is $\mathcal{S} = | \boldsymbol{r_{1}} \times \boldsymbol{r_{2}} |$.
The Fourier coefficients are found by integration over one unit cell of the lattice. With $\boldsymbol{k}_{nm}\equiv n\boldsymbol{K}_1+m\boldsymbol{K}_2$,  
\begin{eqnarray}  
  C_{nm} & = & \frac{1}{I_M \mathcal{S}  } \int_{\mathcal{S} } 
    M_2(\boldsymbol{\rho})\cos[\boldsymbol{k}_{nm}\cdot\boldsymbol{\rho} ]\;d^2\boldsymbol{\rho},\label{Cnm} \\
  S_{nm} & = & \frac{1}{I_M \mathcal{S}  }\int_{\mathcal{S} }
    M_2(\boldsymbol{\rho})\sin[\boldsymbol{k}_{nm}\cdot\boldsymbol{\rho} ]\;d^2\boldsymbol{\rho}. \label{Snm}
\end{eqnarray}
For the  magnetic potential we obtain \cite{hinds1999magnetic}:
\begin{eqnarray} \label{eq4}
  \Phi_M(\boldsymbol{r}) = \frac{1}{2}\mu_0 I_M \sum_{n,m=-\infty}^{\infty}
  e^{-k_{nm}z} [ & & \hspace{-1em} C_{nm} \cos(\boldsymbol{k}_{nm}\cdot\boldsymbol{\rho} ) + \nonumber \\
  & &  \hspace{-1em} S_{nm}   \sin(\boldsymbol{k}_{nm}\cdot\boldsymbol{\rho} )\, ]  
\end{eqnarray}
$C_{nm}$ and $S_{nm}$ now give a compact description of the magnetic field configuration. $ \Phi_M(\boldsymbol{r}) $ is defined such that it is a scale invariant Fourier series. This way $\boldsymbol{B}_{\mathrm{film}}(\boldsymbol{r})=-\nabla \Phi_M(\boldsymbol{r})   \propto a^{-1} $ and  $ \nabla  \boldsymbol{B}(\boldsymbol{r}) \propto a^{-2}$, where $a$ is the lattice spacing. When the field from the magnetic pattern is combined with an external field, magnetic potential wells are created, which can be described by
$$ \boldsymbol{B}_{\rm{tot}}  = \boldsymbol{B}_{\rm{ext}} + \boldsymbol{B}_{\mathrm{film}}.  $$

\begin{figure}[h]  
\centering
\includegraphics[width = 0.9 \linewidth   ]{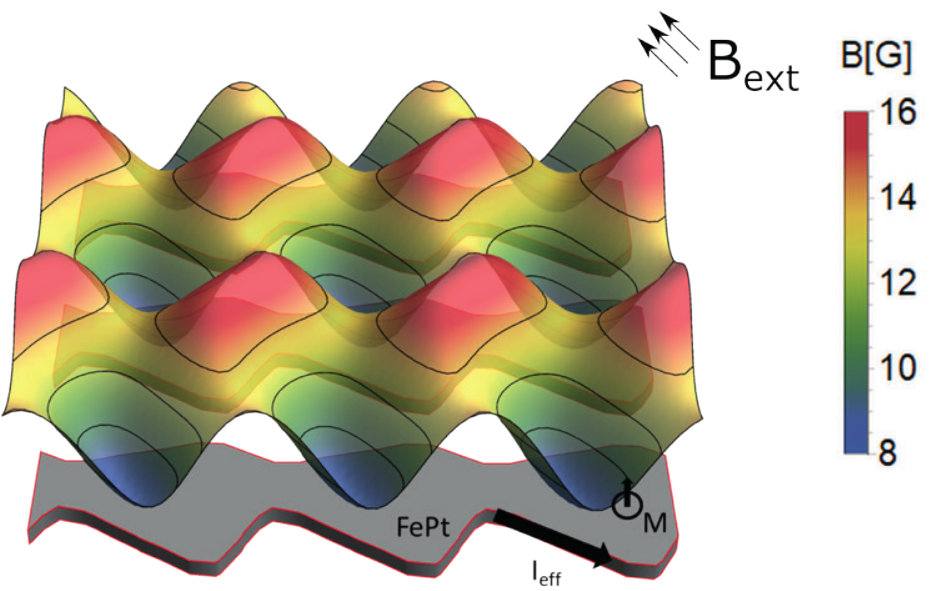}
\caption{Three-dimensional representation of the magnetic potential in arbitrary units above an out-of-plane magnetized patterned layer of FePt (gray). The potential is calculated at a height $z=0.5a$ above the surface, with $a$ the lattice constant. The external magnetic field $\boldsymbol{B}_{\rm{ext}}$ was chosen such that a square trapping potential with equal barrier heights at $z=0.5a$ is obtained: $B_{ext} = $ \unit[(-10, -9.0, -$5.0 \times 10^{-2}$)]{G}.}
\label{superrooster}
\end{figure}

In Fig.\ \ref{superrooster} the magnetic film for the square magnetic lattice is pictured together with the potential created by film and external field. In this case the pattern was calculated by an optimization scheme for p2 (``wallpaper group") type lattices \cite{schmied2010optimized}. All trap parameters depend on the value of the external field and $I_{M}$. Without an external field no traps exist, and depending on the magnitude and sign of its components, traps can be created at different positions above a particular chip pattern. By increasing (lowering) the external field, the  trap frequencies can be raised (lowered).   

\begin{figure}
\centering
\includegraphics[width = 0.9 \linewidth  ]{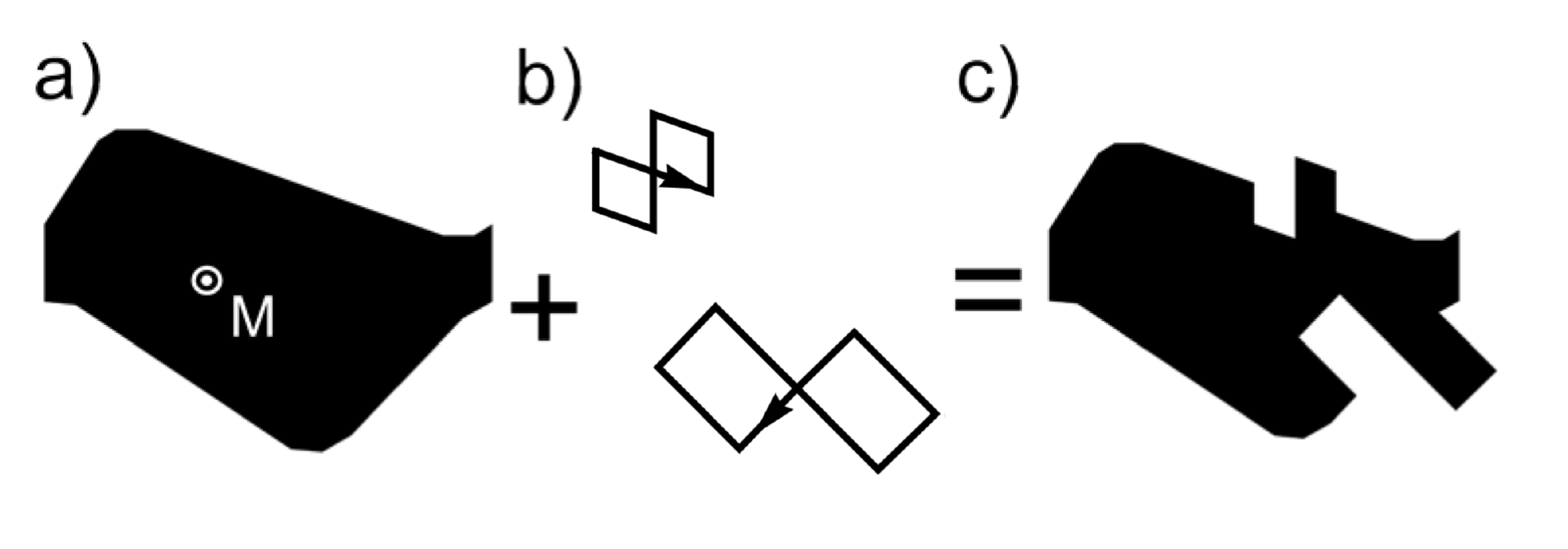}
\caption{  (a)  The tile of the triangular lattice with magnetization $M$ out of plane. (b) Two loop wires that are added to the unit cell, one on each side of the unit cell, with one loop section canceling the equivalent edge current. (c) The resulting modified unit cell.} \label{fig:defect}
\end{figure}

To create lattices of other symmetry classes than this p2 class, we combine a periodic (square or triangular) pattern with designer defects. The edge of the magnetic film is changed locally to raise or lower specific potential minima. We introduce these defects as small virtual loop wires with current $I_{M}$; see Fig.\ \ref{fig:defect}. 
The total field is now given by
\begin{equation}
\boldsymbol{B}_{\rm{tot}}  = \boldsymbol{B}_{\mathrm{ext}} + \boldsymbol{B}_{\mathrm{film}}  + \boldsymbol{B}_{\mathrm{defects}} .
\end{equation}
Calculating the magnetic field potentials like this is more accurate than calculating the sum of a finite number of loop wires. When the whole pattern is described by loop wires, edge effects often overshadow the lattice structures. This can be overcome in principle by taking a large enough number of loop wires. However, this quickly becomes computationally expensive.

In the sections below we calculate the designed potentials. For all presented lattices we calculated the bottom field and the field at the saddle point(s),  which form the potential barriers between the traps in the different lattice directions. We consider lattices where the lattice spacing $a$ is several micrometers, where controlled interactions can be created between trapping sites \cite{urban2009observation, weimer2010rydberg}. For all trapping potentials the trap frequencies can be found by making a harmonic oscillator approximation at the trap minimum \cite{gerritsma2006topological}.  

In this paper we concentrate on lattice constants of a several micrometers. The distances of the traps to the surface will then be at approximately the same length scale, which can be understood as a consequence of the exponent in Eq. (4). Many experiments on the excitation of Rydberg atoms close to a surface have faced the issue of stray electric fields and field gradients, to which Rydberg atoms are very susceptible \cite{tauschinsky2010spatially, naber2016adsorbate}. Such fields and gradients emanate from the surface, usually as a result of atoms adsorbed on the surface. 

Several experiments have demonstrated techniques to reduce these stray fields \cite{davtyan2018controlling, cisternas2017characterizing, sedlacek2012microwave, hermann2014long}. A dramatic reduction was achieved by depositing a thin film of metallic Rb in a cryogenic environment \cite{hermann2014long}. These authors measured a field below \unit[0.1]{V/cm}, and gradients on the order of \unit[1]{V/cm$^2$} at \unit[150]{$\mu$m} distance to the surface, which was  sufficiently small for narrow linewidth Rydberg excitation. For smaller distances little is known so far, but the quoted results are encouraging. We note that the designed patterns presented here are to a large extent scale invariant, so that they remain valid also outside the realm of Rydberg physics.

\section{Triangular lattices with designer defects: kagome and honeycomb} \label{triHoneyKagome}

While the triangular lattice has been used for many quantum lattice studies, nowadays lattices of a higher symmetry class attract more attention. The kagome lattice is of special interest because of its predicted frustrated phases for particles with antiferromagnetic interactions 
\cite{you2012superfluidity, chernyshev2015order, evenbly2010frustrated}. Recent work also predicts the observation of a spin ice phase for Rydberg p-state interactions on a kagome lattice \cite{glaetzle2015designing,carrasquilla2015two}. The kagome lattice can be created by shaken optical lattices \cite{hauke2012non,jo2012ultracold} and with an array of dipole traps \cite{labuhn2016tunable}. Although both techniques have their particular strengths, a magnetic chip based potential could provide a scalable lattice without harmonic confinement and with control over the lattice boundary. 

The similar honeycomb structure has attracted much attention in recent years because of the presence of Dirac cones in its band structure, which gives rise to the many extraordinary properties of graphene. A magnetic hexagonal lattice could be used to perform quantum simulations of graphene or to search for other nontrivial quantum phases which are predicted to arise for hard-core bosons in graphene-like geometries 
\cite{haldane1983nonlinear, varney2012quantum}. Optical realizations of the hexagonal lattice have been either irregular (stretched) \cite{labuhn2016tunable} or spin dependent \cite{flaschner2016experimental, flaschner2018high}, while with a magnetic lattice the exact honeycomb structure may be realized. Using nanofabrication techniques, even an interface between a frustrated kagome lattice and a nonfrustrated hexagonal lattice can be created. 

Rydberg atoms have been successfully used for quantum simulations in lattices created by optical dipole traps of approximately \unit[3]{$\mu$m} spacing \cite{labuhn2016tunable,saffman2010quantum,zeiher2016many}. These experiments can only trap up to approximately 50 atoms while a magnetic lattice can create much larger lattices \cite{leung2014magnetic}.

To build these more exotic lattices, we start from a triangular lattice. With the methods of Ref. \cite{schmied2010optimized} a triangular lattice can be generated with traps in a lattice spanned by the vectors $\boldsymbol{r}_{1} = (1,0) $ and $\boldsymbol{r}_{2} =  ( \frac{1}{2} , \frac{\sqrt{3}}{2} )$. Approximately equal barriers between all the trapping positions can be found by applying the appropriate external field. Note that three different barriers exist, in the directions of $\boldsymbol{r_{1}}$, $\boldsymbol{r_{2}}$, and $\boldsymbol{r}_{3}  = \boldsymbol{r}_{1}-\boldsymbol{r}_{2}$.  Assuming a film thickness of $\unit[50]{nm}$, a magnetization of \unit[800]{kA/m},  and a trap at the height of $z=0.5 a = $ \unit[2.5]{$\mu$m},  one finds a trap depth of \unit[6.6]{G}, a bottom field of   \unit[2.2]{G}, and trap frequencies of \unit[(60, 58, 18)]{kHz}. Larger (smaller) trap frequencies and barriers can be created with thicker (thinner) magnetic films and/or stronger (weaker) external fields. 

\begin{figure} 
\centering
\includegraphics[width=0.99 \linewidth ]{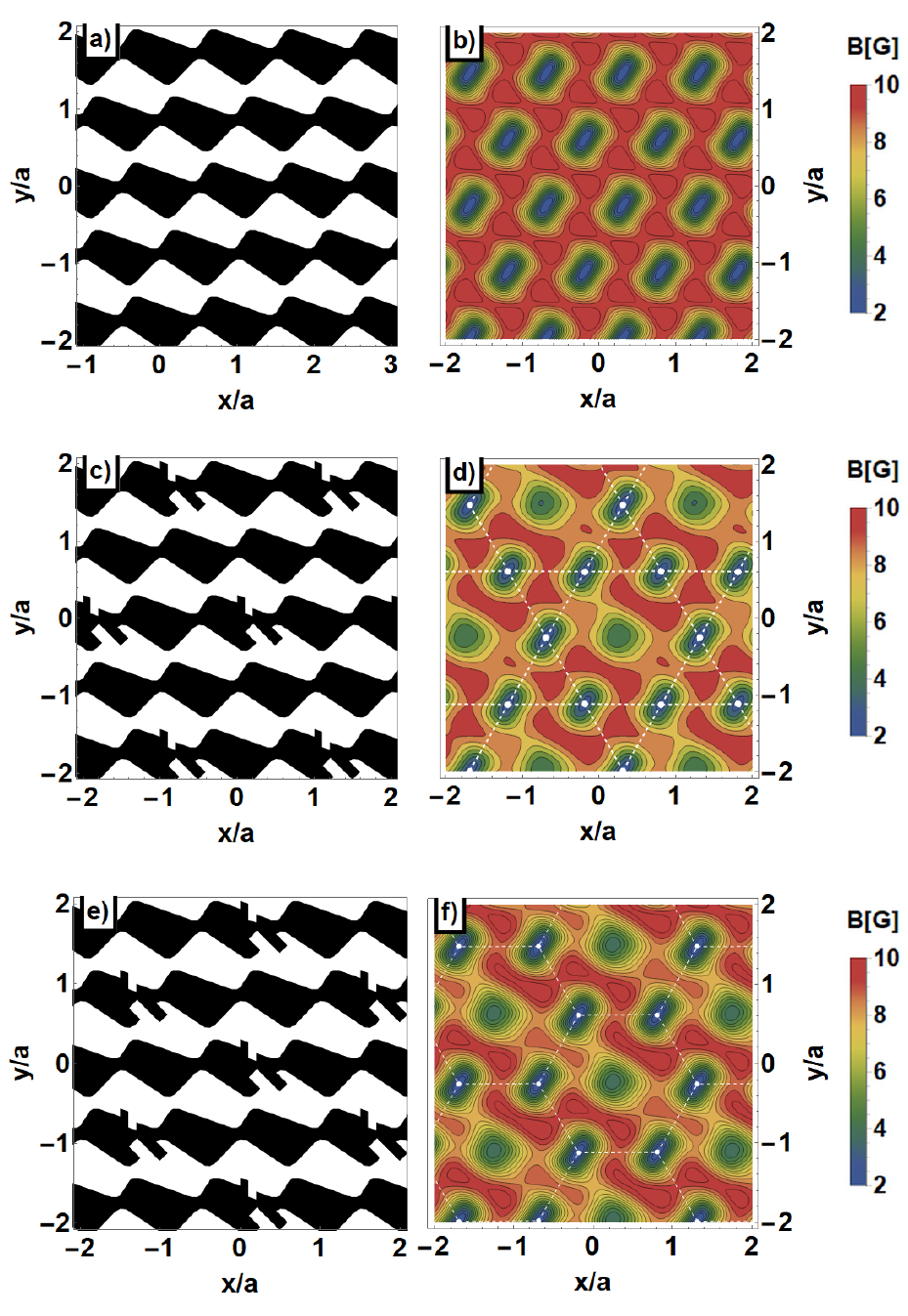}   
\caption{(a) The triangular structure with $a = $ \unit[5]{$\mu$m}: (b) the corresponding potential taken at $z=0.5 a = $ \unit[2.5]{$\mu$m}. The external field is given by $ \boldsymbol{B}_{\rm{ext}} =  $ \unit[( -5.0, 2.0, -0.2)]{G} and the trap bottom field is \unit[2.2]{G}.  (c) The kagome lattice structure. In black  the magnetic pattern is shown with modified local shapes at the trap positions that needed to be raised. (d) The  potential created by pattern (c) with a lattice spacing of  \unit[5]{$\mu$m} and trap minima at $z=0.5a$. The same external magnetic field as (b) is applied. The kagome lattice is indicated by the white dashed lines.   (e) The honeycomb lattice structure. In black  the magnetic pattern is shown with modified local shapes at the trap positions that needed to be raised. (f) The  potential created by pattern (e) with a lattice spacing of  \unit[5]{$\mu$m} and trap minima at $z=0.5a$. The same external magnetic field as in (b) is applied.  The hexagonal lattice is indicated by the white dashed lines.   }  \label{triangular}

\end{figure}

\subsection{The kagome lattice }
Starting from an optimized pattern for a triangular lattice, we can create the kagome lattice by introducing designer defects that raise the potential at specific sites. These are local modifications of the edge of the pattern underneath the default trapping position of the base triangular lattice. These defects can be described by virtual loop currents such that one edge of the loop cancels a section of the original edge.  An example is given in Fig.\ \ref{fig:defect}.  Loops are chosen such that the total magnetic coverage stays 50\%, which is important to avoid local magnetic field gradients within the lattice.  

The kagome lattice is obtained by adding two of these defects to a subset of the triangular lattice sites. A combination of two defects with different size, position, and orientation is used to obtain the desired potential. We consider a realization with a \unit[5]{$\mu$m} lattice spacing that is designed for Rydberg atom experiments. For comparison we calculate the trap parameters for traps centered around $z=0.5 a = $ \unit[2.5]{$\mu$m} with the same external field as for the triangular lattice in Fig.\ \ref{triangular}.  In these lattices, at the sites with the defects the potential wells are raised by  \unit[1.5]{G}. The remaining (nonraised) potential wells then form the desired kagome  sublattice. In Figs.\ \ref{triangular}(c) and \ref{triangular}(d) the kagome lattice structure is presented with potential raising defects. The sublattice period is given by multiples of the vectors $ 2 \boldsymbol{r_{1}} $ and $ 2 \boldsymbol{r_{2}}$.

\subsection{The honeycomb lattice}
The hexagonal lattice is constructed in a similar way. Now the tiles including defects are placed on the sub-lattice that is created by multiples of $\boldsymbol{r_{1}} + \boldsymbol{r_{2}}$  and  $ 2 \boldsymbol{r_1} - \boldsymbol{r_{2}}$; see Figs.\ \ref{triangular}(e) and \ref{triangular}(f). The bottom field of the elevated traps is raised to \unit[3.8]{G} while the lower traps are kept at \unit[2.3]{G}, as for the kagome lattice. This demonstrates the universal applicability of this technique.

\subsection{Fabricated magnetic honeycomb and kagome lattices }
\begin{figure} [htb]
\centering
\includegraphics[width=0.9 \linewidth   ]{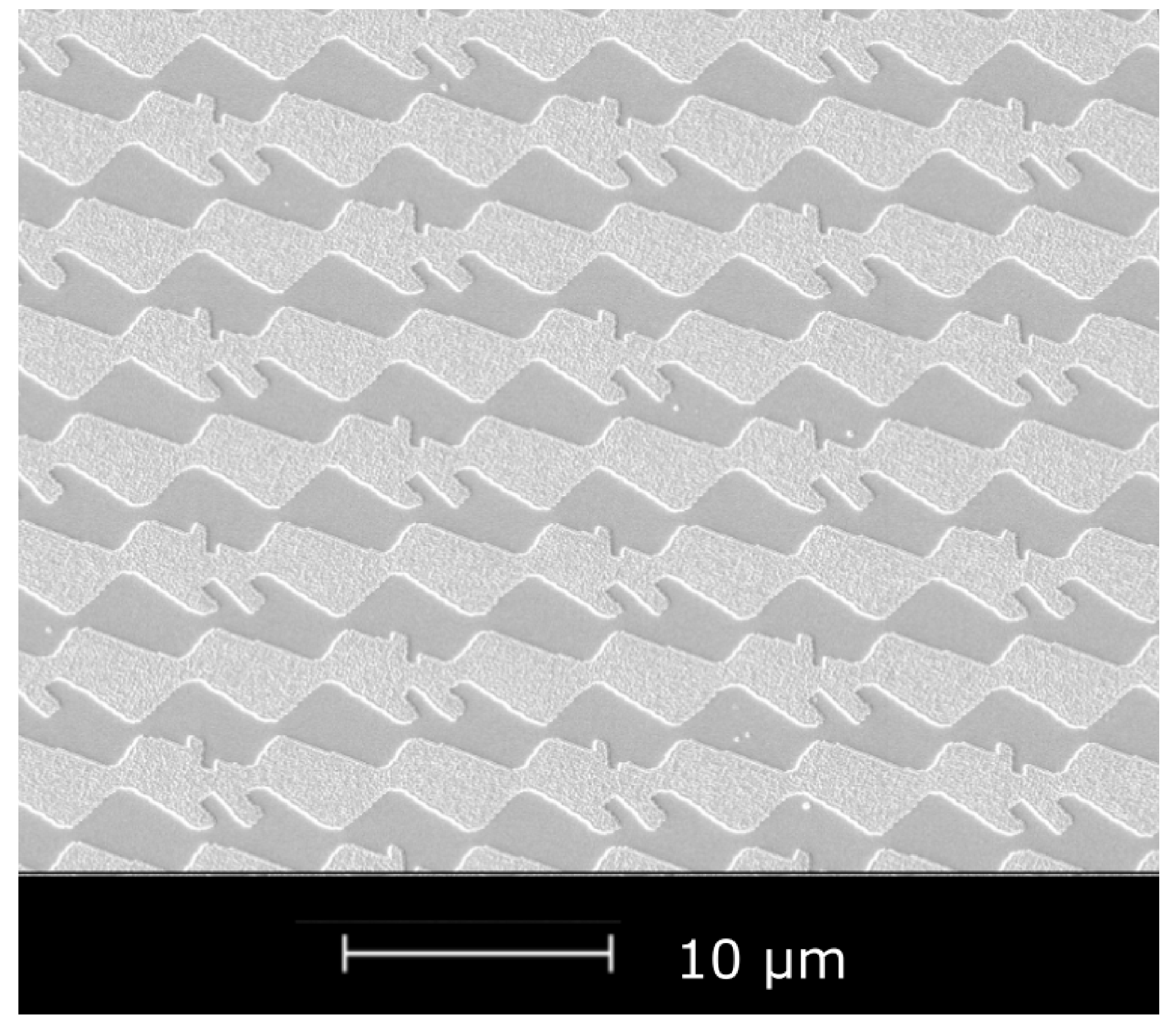}  
\caption{ SEM image of the honeycomb lattice with lattice spacing of \unit[5]{$\mu$m}. Lighter regions indicate magnetic material. }  \label{SEM}
\end{figure}

The kagome and honeycomb structure have been created in a \unit[50]{nm} thick film of magnetized FePt. The fabrication of these magnetic structures is discussed in full detail in a separate paper \cite{la2018deposition}. In Fig. \ref{SEM} we show a scanning electron microscope image of the hexagonal geometry that correspond to the design presented in Figure \ref{triangular}. Figure \ref{SEM} shows several tens of trap sites for the honeycomb lattice. The magnetic material (FePt) scatters more electrons than the substrate (MgO), which makes it appear bright. The structures are covered by a \unit[50]{nm} thick protective Pt layer. 

\section{Low-dimensional structures: ladders and diamond chains} \label{ladders}
We can also employ our magnetic lattices to create low-dimensional structures for ultracold atoms. By local modifications to a lattice potential we can carve out (quasi-)one-dimensional potentials. Low-dimensional structures form a natural testing platform for many quantum simulation experiments and theories \cite{bernien2017probing, barredo2015coherent, endres2016atom,trotzky2012probing}. Here we provide several geometries that can be used to simulate spin models such as those proposed in Ref. \cite{wei2013effective, tschischik2015bose, uchino2015two} and 
\cite{hyrkas2013many, pelegri2018topological, huijse2012supersymmetric}. The simulations of nonperiodic lattices that are presented here are based on finite lattice structure calculations. The magnetic film is described by calculating the magnetic field of the equivalent edge current $I_{M}$ for each pattern. We present only the inner-most region of the total structures that we calculate, since the edge of all structures perturbs the periodic patterns in typically the first ten unit cells from the edge. If the edge is not placed far enough from the shown central region, edge effects can be recognized by an extra magnetic gradient in the direction of the bias field. No such edge effects are present in any of the presented lattices. In the experiments we typically prevent these effects in a similar way by surrounding our lattices with a region several millimeters wide containing other lattices such that we maintain a 50\% magnetic coverage \cite{la2018deposition}.   

\subsection{Ladders}
\begin{figure}
\centering
\includegraphics[width=0.9 \linewidth   ]{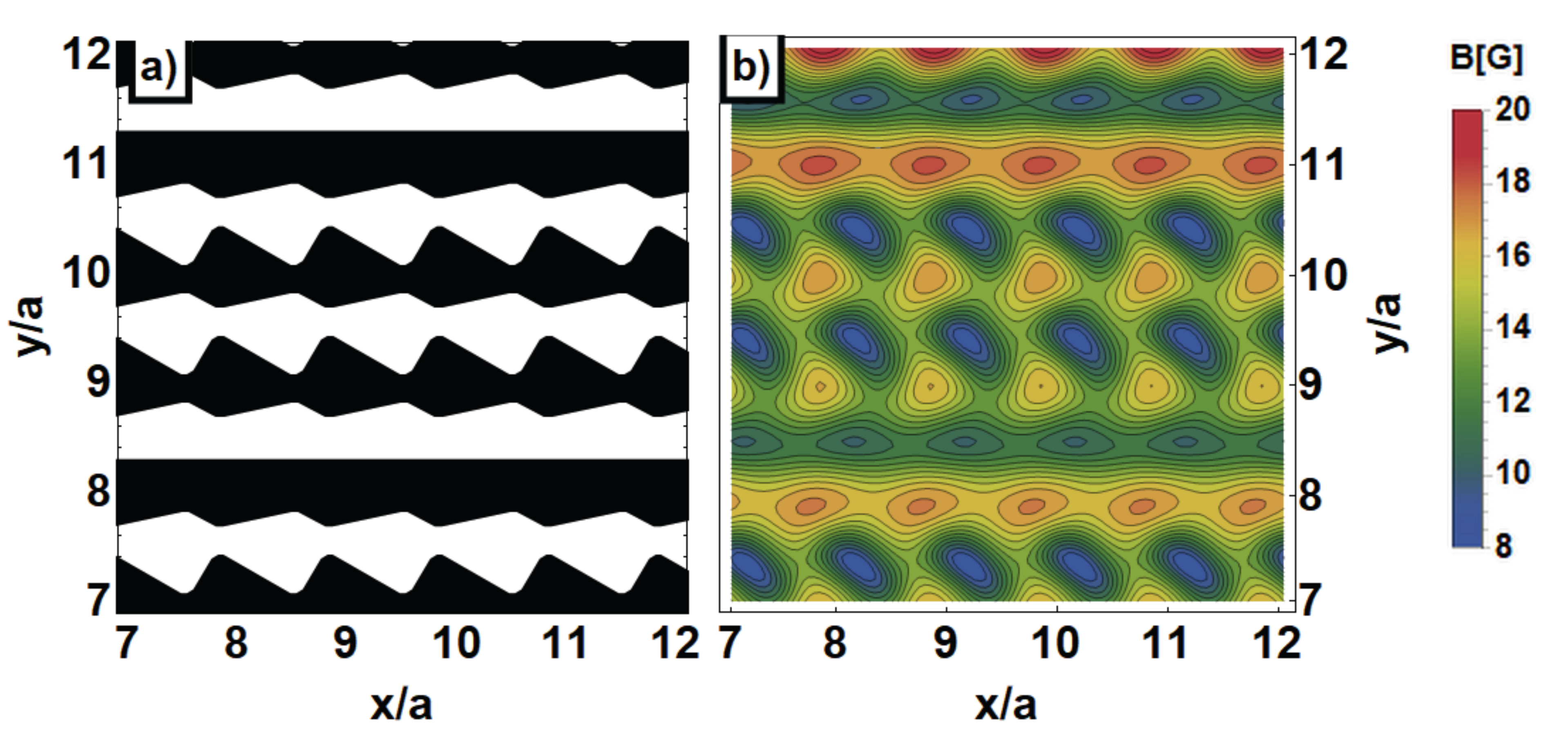} 
\caption{(a) Magnetic film pattern with $a =$ \unit[5]{$\mu$m} that can create a double-well series or two-rail ladder. (b) The external field is here set to generate equal barriers between all ladder sites and is $ \boldsymbol{B}_{\rm{ext}} =  $ \unit[( -10, -9.0, $-5.0 \times 10^{-2}$)]{G}. }\label{Doublewells_fig5}
\end{figure}

To construct confining barriers in two-dimensional lattice potentials, we seek a method to separate traps into particular arrays. Small spin chains or two-dimensional plaquettes of $ 2 \times 2 $ traps can be produced by isolating sets of traps. Inspired by optical lattice techniques where local traps are raised by overlapping optical fields, we designed magnetic artifacts that create a sharp magnetic barrier, similar to the defects presented in the previous section. By straightening out a horizontal magnetic array  we were able to disrupt the lattice in the $y$ direction. These interrupting ``fences" can be used in combination with double or triple arrays of square lattice traps to construct ladders of a chosen number of rails. 
To construct a one-rail ladder, or equivalent a series of double wells, we alternate two trapping arrays with one fence array. Here we chose the external field such that equal barriers in the $x$ and $y$ directions are obtained between the traps in the ladder. With the direction and magnitude of $B_{\mathrm{ext}}$ we can control the position and confinement of the traps. For example it is possible to produce a one-dimensional lattice of double wells with the same magnetic structure.  In Fig. \ref{Doublewells_fig5} the corresponding potentials are presented.  When considering bosons in ladder geometries like this the Haldane model predicts a gapped phase for odd-rail ladders, while for even-rail ladders the energy bands touch \cite{haldane1983nonlinear,martin1996phase}.

\begin{figure}
\centering
\includegraphics[width=0.9 \linewidth   ]{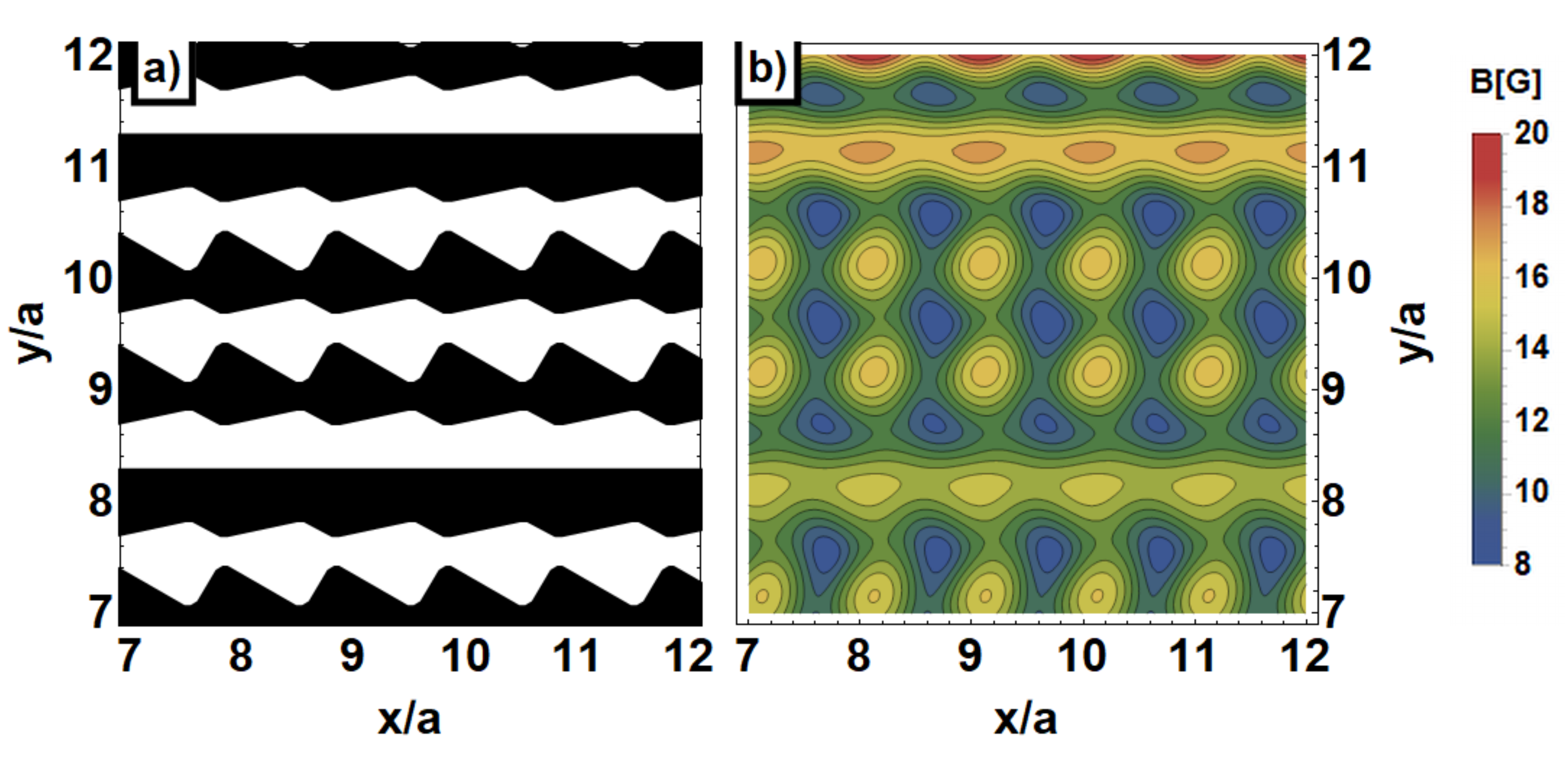} 
\caption{(a) The same magnetic film pattern as Fig.\ \ref{Doublewells_fig5}. (b) The potential created by the film of (a) with the reversed external field: $ \boldsymbol{B}_{\rm{ext}} =  $ \unit[(10, 9.0, $5.0 \times 10^{-2}$)]{G}. A three-rail ladder is created by the same structure as Fig.\ \ref{Doublewells_fig5} with reversed external field to create wells above the other edge of the magnetic structures.  }\label{Doublewells_fig6}
\end{figure}

A three-rail ladder can similarly  be created by using either three trapping ladders, as a trivial extension of the previous figure, or by trapping on the other edge of the same magnetic structures by reversing the external field. This is possible because of the periodic structures that are present on the nonflattened side of the barrier arrays. An example can be seen in Fig.\ \ref{Doublewells_fig6}.

\subsection{ Plaquettes }
The horizontal barriers from the previous section can be combined with vertical barriers to create plaquettes of arbitrary sizes. With these structures one is be able to control the number of interacting atoms and the size of their surrounding lattice. Plaquettes may be useful for quantum information applications with trapped atoms in order to implement error correction \cite{weimer2010rydberg,weimer2011digital,saffman2016quantum}. With a magnetic plaquette potential one can employ the scalability and technical benefits of an atom chip to create a controlled grid of atomic traps.  

\begin{figure}
\centering
\includegraphics[width=0.99 \linewidth   ]{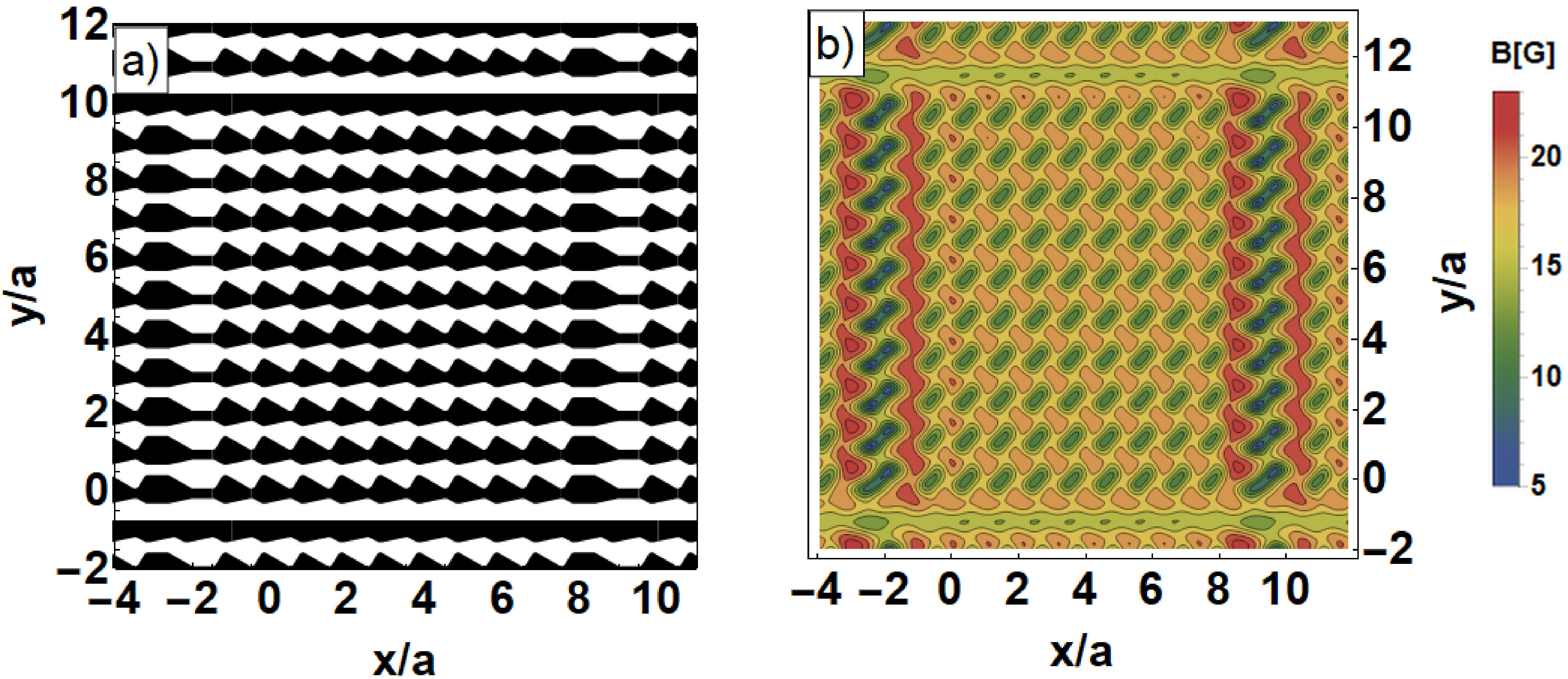} 
\caption{(a) Magnetic film pattern with $a =$ \unit[5]{$\mu$m} where horizontal and vertical barriers have been combined to isolate a plaquette of $10 \times 9$ sites. (b) The potential created by the film of (a) with the same external field as for a square lattice with periodic barriers:
$ \boldsymbol{B}_{\rm{ext}} =  $ \unit[(-10, -9.0, $-5.0 \times 10^{-2}$)]{G}.}\label{plaquette}
\end{figure}

To create the vertical barriers, the corrugation from the top of one magnetic structure was removed. For the horizontal barrier a similar technique is used. By straightening our a single unit cell, a raised barrier is created on its right side together with a low valley on its left. Therefore we need to combine two elements to create a full barrier in the horizontal direction, one straightened-out thin element to create a barrier on the right, and a thick piece of magnetic material to the left of this to isolate the plaquettes from the lower valley. In Fig. \ref{plaquette} one can see how this combination of a thick and a thin element is used to raise a barrier around a plaquette.

\subsection{Diamond chains}
Another geometry of interest for many quantum simulation experiments is the diamond chain. The alternation between the number of sites in each column has made the diamond chain an inspiring tool for spin model proposals \cite{hyrkas2013many,pelegri2018topological,huijse2012supersymmetric,takano1996ground}. For antiferromagnetic spin interactions, several ideas  have been  proposed which show how frustration in complex geometries can lead to new phases of matter \cite{canova2009exact, schmidt2008supersolid}. Due to its presence in many crystalline materials it is also a highly relevant geometry for quantum simulation. Both materials with oxide planes \cite{kikuchi2005experimental,morita2017static, johnston2016electron} and azurite \cite{honecker2011dynamic,rule2011dynamics} have been studied intensively theoretically, but so far have not been realized with optical lattices \cite{kobayashi2016superconductivity,ananikian2012thermal,murmann2015antiferromagnetic}. This is due to the large number of wavelengths that one would need to combine. Their barrier heights vary in different lattice directions and modulate along the chain axes, which makes them hard to emulate with optical techniques. This complexity make diamond chains ideal systems to be studied with magnetic lattices because their complex geometry can be easily handled with nanofabrication techniques.  
 
\begin{figure}
\centering
\includegraphics[width=0.9 \linewidth   ]{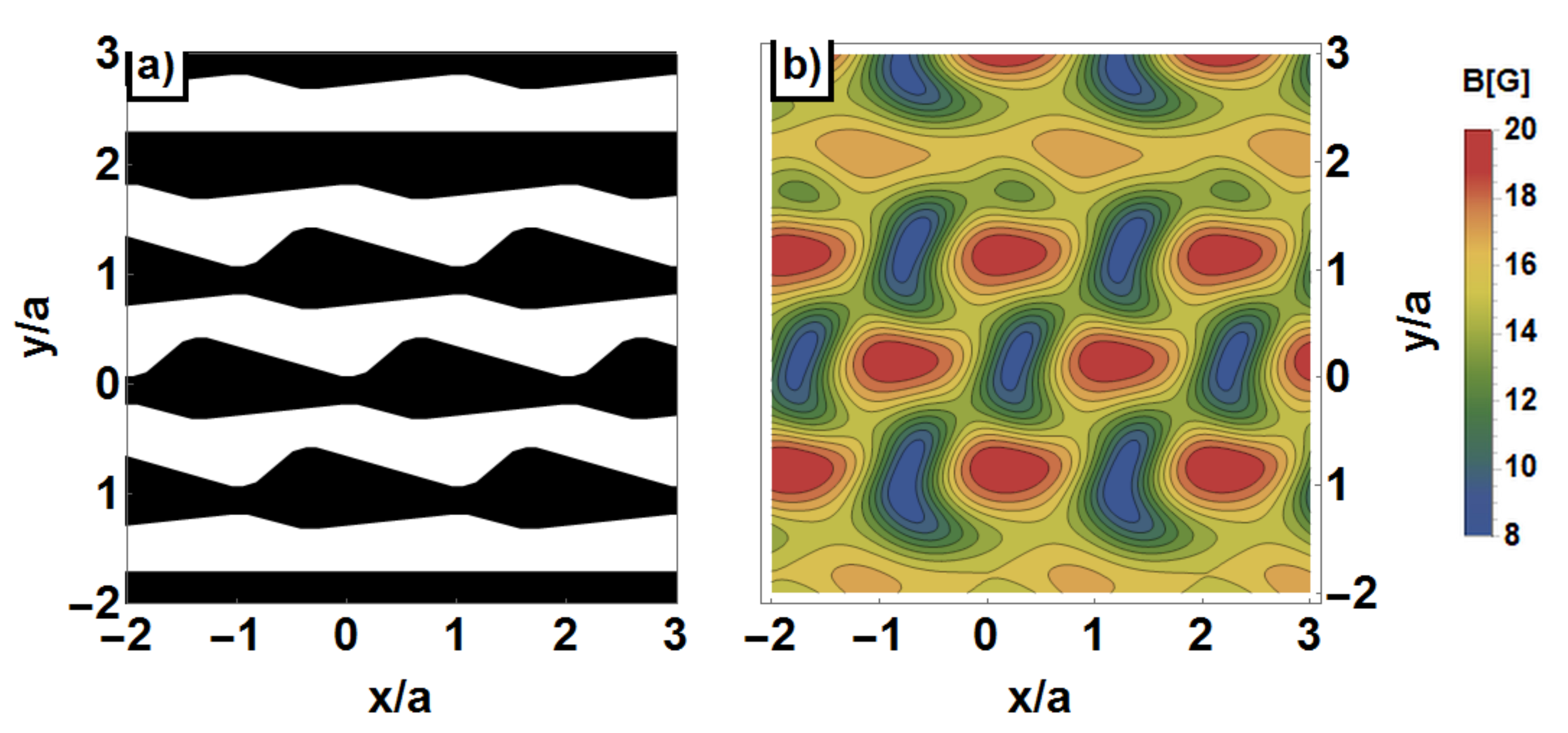} 
\caption{(a) A single diamond chain where single traps are positioned in between double wells. (b) Potential corresponding to the pattern of (a) with an external field $ \boldsymbol{B}_{\rm{ext}} =  $ \unit[(24.4, 2.05, 0.13  )]{G} chosen such that traps are formed at $z= 0.45 a $. Here $a = $ \unit[5]{$\mu$m}.  By controlling the external field, trap barriers in different directions can be controlled independently.  }
\label{diamond3}
\end{figure}

By combining linear stretches and shifts it is possible to create  diamond chain potentials. An example is the single diamond chain shown in Fig.\ \ref{diamond3}. Higher order chains can be created trivially by increasing the number of trapping arrays in between the blocking arrays.

\section{Tapered structures} \label{tapersec}
\begin{figure} 
\centering
\includegraphics[width=0.8 \linewidth   ]{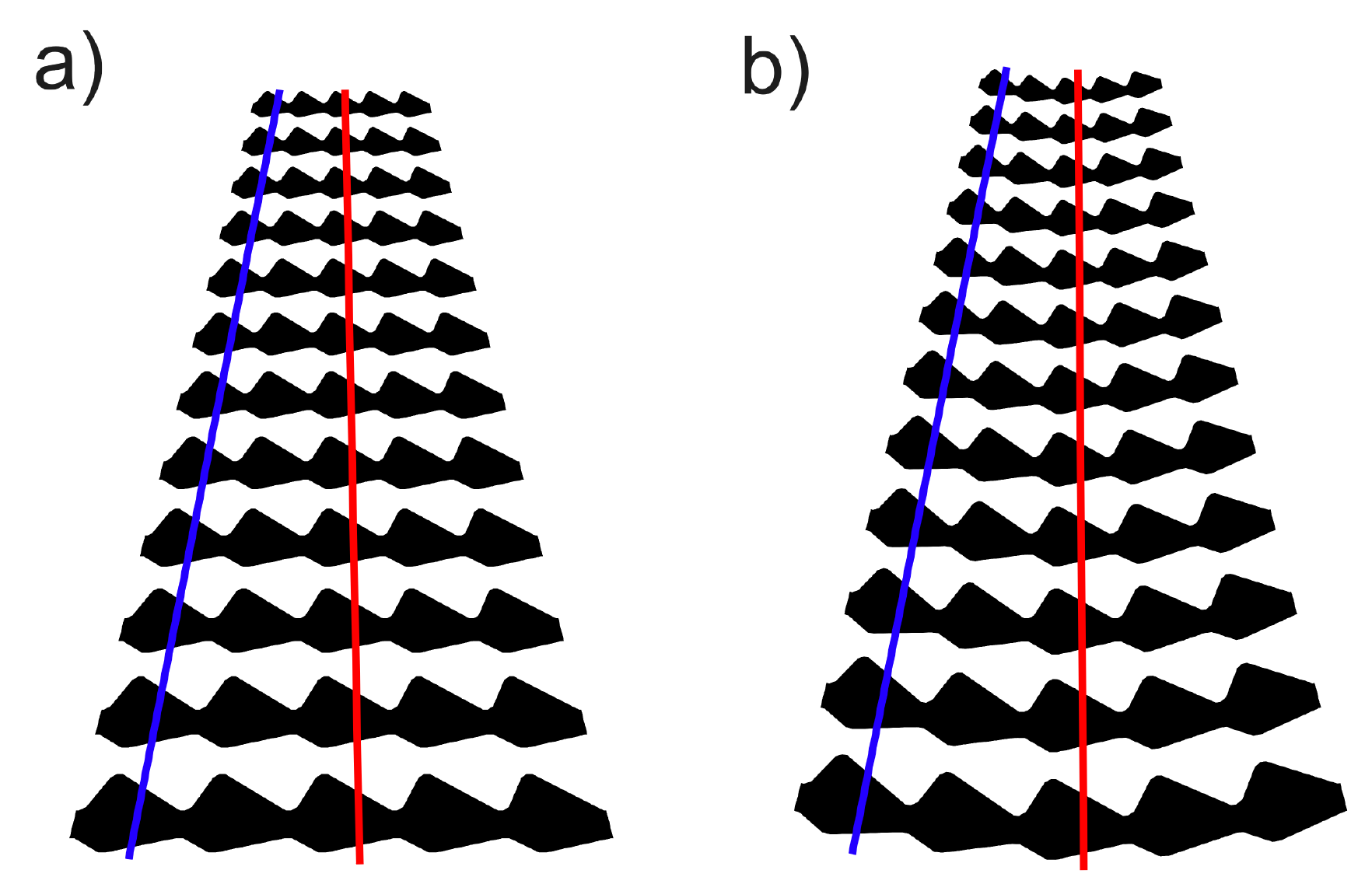}  
\caption{ (a) Example of a straight taper of 11 lines shrinking by a factor of 2 (5\% per line). (b) Example of a rounded taper. The unit cells are now also rotated such that the lattice locally keeps its original shape. The traps along the two red lines are similar but the traps along the blue lines differ due to the different transformation.  } \label{Taper}
\end{figure}

In this section we introduce tapered lattices: structures where the lattice spacing changes gradually across the lattice; see Fig.\ \ref{Taper}. We will consider tapers that can connect the Rydberg regime that we have considered so far to submicron lattices of \unit[250]{nm}. Tapered lattices solve two problems: transport of atoms into submicron lattices and detection of atoms in submicron lattices.
By several groups an effort is being made to reduce the period of atomic lattices, since the interaction energy between atoms as well as hopping rates between sites scales quadratically with the inverse lattice spacing. Loading atoms into these traps from a macroscopic ultracold cloud is hard and inefficient 
\cite{wang2017trapping, parsons2015site, goban2014atom, naides2013trapping, long2005long, gupta2007cavity, greiner2001magnetic}. The loading efficiency gets worse when one shrinks down a lattice. However, it is possible to move all trap arrays from row to row in any of the presented lattices by rotating the external magnetic field around the Ioffe axis periodically \cite{whitlock2009two}. By loading atoms into a large lattice before bringing them closer together using the taper, extra atomic losses could potentially be reduced.  If we transport back through the tapered structure to the larger lattice spacings, we are able to move clouds from a subwavelength lattice up to a region where individual traps can be distinguished by a microscope.

From a more fundamental point of view one might also consider a cloud of atoms trapped in a subwavelength lattice at the smallest end of a tapered structure, which is then released or shifted into the widening taper. Such systems then provide a very natural quantum simulation of a quantum gas in an expanding lattice. One thinks of cosmological theories in which particles in the early universe are considered on an inflating lattice. Experiments like this have also been proposed in the  context of the Kibble-Zurek mechanism (KZM) \cite{kibble1976topology,zurek1996cosmological}, which predicts the formation of domains after a homogeneous gas is released in such a geometry. Although some experiments have been done in periodic optical lattices \cite{schreiber2015observation} and in one-dimensional clouds  \cite{gring2012relaxation}, a two-dimensional lattice experiment with varying length scales within the lattice as proposed here has not been created. 

To combine lattices of different length scales, we developed tapered lattices in which the lattice spacing is varied slowly in one direction.  In optical lattice systems this would require one to vary the frequency of all lattice beams in time over a wide range, while with lithographic patterning one is completely free in the scope and gradient of the lattice spacings.

The combination of the shift array with the shrinking lattices will therefore allow us to capture mesoscopic clouds of hundreds of atoms in traps several micrometer apart and then move them to smaller geometries. One has to limit the amount of change from one unit cell to the next such that similar traps are created in neighboring rows. This way, atoms can be transported adiabatically up and down the lattice.
There are different ways to create tapers, as can be seen in Fig.\ \ref{Taper}. In Fig.\ \ref{Taper}(a) the straight tapered lattice shows deformation at the edge of the taper. Another choice is rounded tapers that are created by a local transformation and rotation of the unit cell to keep the local pattern constant, as shown in Fig.\ \ref{Taper}(b). While the traps along the vertical red lines have the same ratio of trap frequencies, the traps at the edges of the structures, along the blue lines, vary in this regard. Because we can only apply a single, uniform bias field that is optimal for one specific trap site because of the local Ioffe axis, traps with a different axis are deformed. A detailed comparison between these geometries and how they influence the transport efficiency of ultracold atoms requires further study.  

If one limits the row-to-row change to 1\%, to ensure adiabatic transport, it requires 300 shifting operations to shrink the lattice spacing by a factor of 20: $ 0.99^{300} = 0.05$. In such a taper the fields required at the large \unit[5]{$\mu$m} scale are approximately \unit[7]{G} to trap atoms at half the lattice spacing. By linearly increasing the external field while moving the traps down the taper, the final trapping field  reaches \unit[203]{G} to trap atoms in a \unit[250]{nm} lattice, \unit[125]{nm} above the surface of the magnetic film.

This small distance raises the question of whether the Casimir-Polder interaction may perturb the trapping potential and thereby shift the trap minimum position, affect trap depth and frequency, or even induce tunneling into the surface. We also note that the chip is covered by a \unit[50]{nm} layer of Pt, so that the nearest surface is only \unit[75]{nm} from the atoms. A calculation in Ref. \cite{leung2011microtrap} showed that the van der Waals interaction ($-C_3/z^3$) does lower the energy barrier towards the surface, but also showed that the resulting tunneling rate remains vanishingly small, even for distances down to \unit[100]{nm} from the surface. Here we consider the Casimir-Polder modification of the van der Waals potential for the covering Pt layer: 

\begin{equation}
U_\text{CP} = - \frac{C_4}{z^3\left(z+3\lambda/2\pi^2\right)},
\end{equation}
where we take $C_{4} = $ \unit[ $ 1.7 \times 10^{-55}$ ]{  Jm$^{4}$ } and $\lambda =$ \unit[785]{nm} for $^{87}$Rb \cite{pasquini2004quantum}. We find that this lowers the total potential at the trap position, $z_0=$ \unit[75]{nm},  by the equivalent of \unit[3.8]{G}. The attractive CP potential thus slightly increases the trap depth relative to the magnetic field far from the surface.   In principle, the  gradient and curvature of $U_\text{CP}$ would change the trap position and trap frequency, respectively. However, for the tight traps at the narrow end of the taper, we find that these changes are negligible. 

\begin{figure} [htb]
\centering
\includegraphics[width=0.99 \linewidth   ]{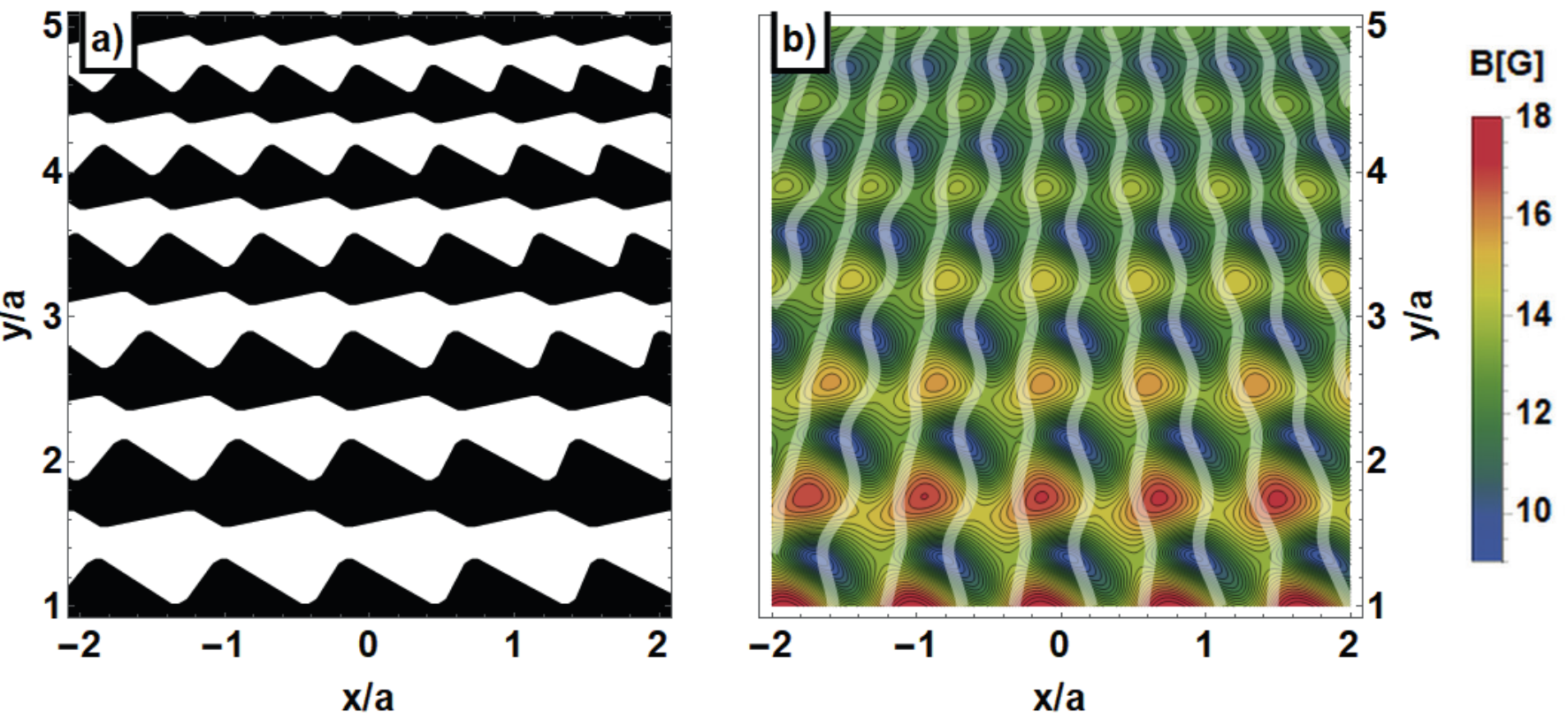}  
\caption{(a) Section of the tapered lattice structure for which the potential is presented. The taper has a slope of 5\% per line and the largest lattice spacing is \unit[5]{$\mu$m}. (b) The potential cross section taken at height $z =0.35 a $ such that it cuts through the trap bottom of the horizontal array at $y=2a$. The white lines show points where possible trap positions can be created. Using a time-varying external field, atoms can be moved along those lines. The external field is $ \boldsymbol{B}_{\rm{ext}} =  $ \unit[(-11, -4.2, 0.20  )]{G}. } \label{Taper2}
\end{figure}

During transport through the taper, the trap frequencies and trap depths increase from  \unit[60]{kHz} and \unit[7]{G} to \unit[17]{MHz} and \unit[203]{G}. In Fig.\ \ref{Taper2} the potential of a tapered section is presented. 
Figure \ref{Taper2}(b) also shows the lines along which transport is possible. The lines have been drawn in plane at height $z=0.35 a$ (for the value of $a$ corresponding to the large end of the taper). These paths can be found by calculating $ \det{ ( \boldsymbol{\nabla}  \boldsymbol{B} ) } = 0 $, which gives all points where a nonzero magnetic field minimum can be created \cite{gerritsma2006topological}.

\section{Conclusion}
We have presented several geometries that can be used in quantum simulation experiments based on permanent magnetic atom chips. With these patterns it has become possible to create magnetic lattices of kagome and honeycomb geometry as well as low-dimensional structures such as ladders and diamond chains. We have shown how some of these geometries have already been realized into a magnetic film. Also magnetic fences have been presented that can isolate plaquettes of atomic traps to lattices with fixed dimensions. Furthermore tapered structures have been created that allow for new experiments in varying geometries and may be used to load subwavelength lattices. It would be interesting to investigate whether the potentials which are presented here on the micron scale can be scaled down to the subwavelength regime. \\

\section{Acknowledgments} 
This work has been supported by the Netherlands Organization for Scientific Research (NWO). We also acknowledge support by the EU H2020 FET Proactive project RySQ (640378). We would like to thank Lara Torralbo-Campo, Jean-Sebastien Caux, Vanessa Leung, M. A. Mart\'{i}n-Delgado, and Vladimir Gritsev for helpful discussions.

\bibliographystyle{apsrev} 
\bibliography{thesisBIBdef_jan2019}

\end{document}